\documentclass[aps,preprint,superscriptaddress,groupedaddress]{revtex4-1}

\usepackage{amssymb,amsfonts,amsmath}
\usepackage{graphicx}  
\usepackage{dcolumn}   
\usepackage{bm}        
\usepackage{epsfig}
\usepackage[T1]{fontenc}
\usepackage{appendix}
\usepackage{color}
\usepackage{cleveref}
\usepackage[utf8]{inputenc}
\usepackage{textcomp}
\usepackage[]{graphicx}
\newcommand{\jk}[1]{\textcolor{black}{#1}}

\begin{document}
\title{Non-linear photochemical pathways in laser induced atmospheric
aerosol formation}

\author{Denis Mongin$^1$}
\author{Jay G. Slowik$^2$}
\author{Elise Schubert$^1$}
\author{Jean-Gabriel Brisset$^{1,3}$}
\author{Nicolas Berti$^1$}
\author{Michel Moret$^1$}
\author{Andr\'e S. H. Pr\'ev\^ot$^2$}
\author{Urs Baltensperger$^2$}
\author{J\'er\^ome Kasparian$^4$}\email{jerome.kasparian@unige.ch}
\author{Jean-Pierre Wolf$^1$}

\affiliation{$^{1}$Universit\'e de Gen\`eve, GAP-Biophotonics, Chemin de Pinchat 22, CH-1211
Geneva 4, Switzerland}
\affiliation{$^{2}$Paul Scherrer Institute, Laboratory of Atmospheric Chemistry, CH-5232
Villigen, Switzerland}
\affiliation{$^{3}$Max-Born-Institut, Max-Born-Strasse 2A, 12489 Berlin, Germany}
\affiliation{$^{4}$Universit\'e de Gen\`eve, GAP-Non-linear, Chemin de Pinchat 22, CH-1211
Geneva 4, Switzerland}

\date{\today}

\keywords{femtosecond laser, filament, Aerosol, condensation, Aerosol Mass Spectrometer}

\begin{abstract}
We measured the chemical composition and the size distribution
of aerosols generated by femtosecond-Terawatt  laser pulses in the atmosphere using an aerosol mass spectrometer (AMS). 
We show that nitric acid condenses in the form of ammonium nitrate, and that oxidized volatile organics also contribute to particle growth. These two components account for two thirds and one third, respectively, of the dry laser-condensed mass. They appear in two different modes centred at 380 nm and 150 nm. The number concentration of particles between 25 and 300 nm increases by a factor of 15.
Pre-existing water droplets strongly increase the oxidative properties of the laser-activated atmosphere, substantially enhancing the condensation of organics under laser illumination.
\end{abstract}


\maketitle
%
%
%
%
%
%
%
%
\thispagestyle{empty}

\section*{Introduction}

Self guided filamentation \cite{kelley_self-focusing_1965,braun_self-channeling_1995,couairon_femtosecond_2007,berge_ultrashort_2007,chin_propagation_2005}
of high-power femtosecond laser pulses results from a dynamic balance between
the Kerr effect which tends to self-focus the laser beam on one side,
and the plasma created by the strong electric field, as well as
higher-order polarisation saturation effects \cite{bejot_higher-order_2010,bejot_transition_2011}
 on the other. The resulting
filaments can propagate
beyond 100 m in the atmosphere \cite{la_fontaine_filamentation_1999},
be created remotely \cite{rodriguez_kilometer-range_2004}, and withstand
adverse atmospheric conditions like clouds \cite{mejean_multifilamentation_2005}
or turbulence \cite{chin_filamentation_2002,salame_propagation_2007}.
This non-linear propagation regime is of particular interest for
atmospheric applications \cite{kasparian_physics_2008} such as remote
sensing \cite{kasparian_white-light_2003}, \jk{air lasing~\cite{Dogariu2011,Yao2011},} triggering of high
voltage discharges \cite{zhao_femtosecond_1995,pepin_triggering_2001,rodriguez_megavolt_2002}
and lightning control \cite{kasparian_electric_2008}. The most striking
effect of laser filaments in the atmosphere is laser-induced condensation
\cite{rohwetter_laser-induced_2010,henin_field_2011,kasparian_laser-assisted_2012,leisner_laser-induced_2013,SunLWJWWGLCLX2013,Ju2014}, where the plasma photochemistry
induced by laser filaments leads to nucleation
of new particles \cite{saathoff_laser_2013} and their subsequent growth \cite{henin_field_2011,rohwetter_modelling_2011}. 

Both laboratory and field studies have shown that laser-induced condensation
is partially associated with the production of locally high ozone and NO$_{x}$
concentrations in the gas phase. Furthermore, ion chromatography of
laser-produced particles sampled on filters has shown substantial
amounts of NO$_{3}^{-}$ ions in the laser-condensed particles \cite{petit_production_2010}.
The formation of highly hygroscopic HNO$_{3}$ has therefore been identified as one major pathway to support laser-induced condensation of aerosol mass, for relative humidity (RH) in atmosphere higher than 70\% \cite{henin_field_2011}. Moreover, ppb-range
traces of atmospheric SO$_2$ have also been shown to increase laser-induced
nucleation of new particles increasing the number concentration by typically one order of magnitude
but without significant total aerosol mass increase. Conversely, laboratory experiments in a cloud chamber showed that ppb-range concentrations of volatile
organic compounds (VOCs) lead to an increase in the concentration of laser-induced particles for sizes up to some hundreds nanometers,
together with an increase of the condensed mass \cite{saathoff_laser_2013}.
This suggests that they both contribute at least to the initial phases
of the nucleation but that mostly organic compounds condense on bigger particles.
However, very little was known to date about the actual composition of
the laser-induced aerosol mass, which is needed to identify the physico-chemical pathways.

Here, we present a detailed chemical characterization of laser-induced aerosols in the real atmosphere. 
An aerosol mass spectrometer is used to investigate the influence
of laser filamentation on ambient particle composition and size distributions. 
We show that condensation indeed occurs in the atmosphere under illumination by laser filaments, leading to an increase of the total aerosol mass and the appearance of a new size mode.
Furthermore, we demonstrate the key contributions of ammonium nitrate (rather than nitric acid, as previously expected), and to a lesser extent, of organics, to the laser-condensed mass. 
Finally, we show that continuously spraying pre-existing water droplets into the laser beam strongly increases the oxidative properties of the laser-activated atmosphere, substantially enhancing the condensation of organics under laser illumination.

\begin{figure}
\centerline{\includegraphics[width=0.95\columnwidth]{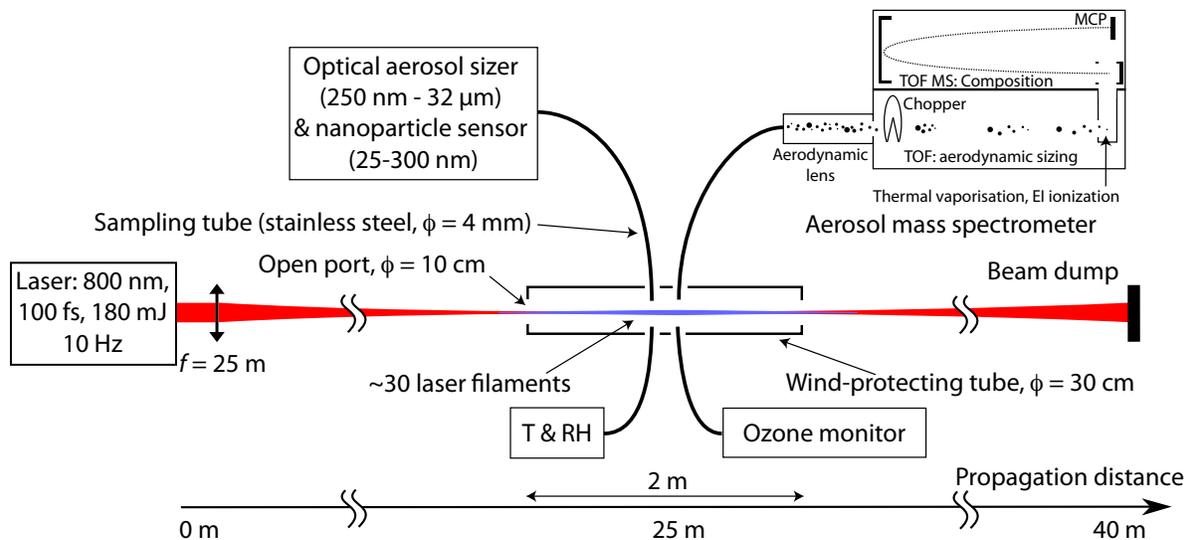}}
\caption{Experimental setup. \jk{TOF: Time of flight; MS: Mass spectrometer; EI: Electron impact; MCP: Micro-channel plate.} \label{fig:experimentalSetup} }
\end{figure}

The experimental setup is sketched in Figure \ref{fig:experimentalSetup}.
The Teramobile \cite{wille_teramobile:_2002} Ti:sapphire laser delivers
Fourier limited, 180~mJ and 80~fs pulses centered at 800~nm at a repetition rate of 10~Hz. The pulses are loosely focused by an $f=25$~m telescope into the atmosphere.
At the focus, the beam  has a waist of around 5~mm (HWHM) and contains $\sim$ 30 filaments of about 100 $\mu$m in diameter each.
The focal region is shielded from the wind by an open, galvanized iron tube of 30 cm diameter.

The atmosphere is  monitored in the filamenting region at 1~cm side distance
from the laser beam by an Aerodyne high-resolution time-of-flight aerosol mass spectrometer (AMS), an optical aerosol sizer, a nanoparticle sensor, and an ozone detector (see Materials and Methods). Additionally, temperature and RH were continuously monitored at the sampling location of these instruments.
Unless otherwise specified, the relative humidity was between 80~\% and 100~\% during the measurements.

Each experiment consists of a series of 2 to 5 cycles of reference
(laser off) and active (laser on) periods of about 15~minutes. The
shielding tube is flushed with fresh air from the open atmosphere before
each laser-on time period.
Besides the nominal conditions described above, we investigated the effect of spraying water droplets in the experimental region using a nebuliser.

We aim at comparing the composition of the pre-existing atmospheric
aerosols with those formed under the influence of
the laser filaments. We therefore focused on steady-state situations,
by averaging the measured size distributions and species concentrations
over the active and reference periods. We corrected the measurement baseline for the slow drift of temperature and RH during the experiments. To avoid interference from
the transient behaviours of both the instruments and the atmosphere just after starting or stopping the laser, we discarded the time interval of $\Delta t=\mathrm{4}$~min at the beginning
of each period. This ensures
that the average over the remaining time period is stable
within 3 standard deviations of the mean. When needed, the effect
of the laser was assessed by performing a Student test for the comparison
of results from the active and reference periods. Significance is
considered achieved for confidence levels of 99\% or beyond ($\alpha \le 0.01$).

\section*{Results}

\subsection*{Aerosol mass increase}

\begin{figure}
\centerline{\includegraphics[width=0.8\columnwidth]{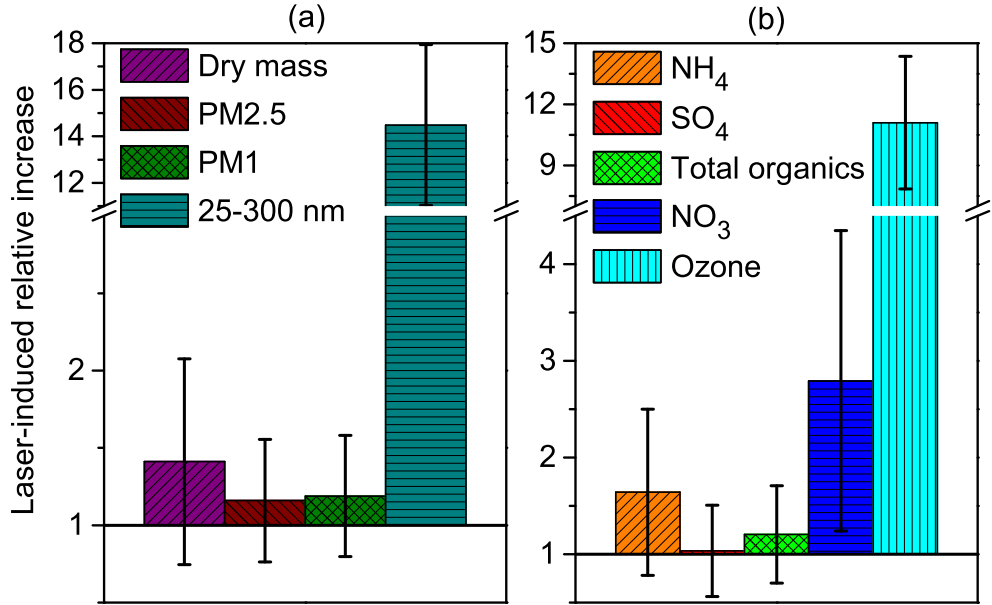}}
\caption{Laser-induced relative increase of (a) the aerosol dry mass concentration measured by the AMS (sum of sulfate, nitrate, organics, ammonium and chloride for vacuum aerodynamic diameters between 60 and 600~nm), the PM1 and PM2 mass concentration measured by the optical sizer and the number concentration of 25 to 300 nnm particles detected by the nanoparticle sensor, and (b) the mass concentration of several condensable components detected by the AMS, as well as atmospheric ozone concentration. \jk{Error bars correspond to the sum of the relative standard deviations of the values measured with and without laser.}\label{fig:mass1}
} 
\end{figure}

The presence of the laser filaments results
in a clear increase of the total aerosol mass concentration. 
As displayed
in Figure~\ref{fig:mass1}(a), this increase reaches a factor of 14
in the case of the 25~--~300~nm particles ($\sim$80~nm median
diameter). Simultaneously, the concentration of  particulate matter with an optical diameter of 250 nm to 1~$\mu$m or 2.5 $\mu$m (PM1 and PM2.5, respectively) increases by almost 20\%, while the AMS measures a 40\% increase of the total dry mass (sum of sulfate, nitrate, organics, ammonium and chloride) for aerodynamic diameters between 60 and 600~nm.
Such values are in line with previous measurements in the atmosphere \cite{henin_field_2011}.
This rise is quite evenly spread over the size classes ranging from
300 to 600~nm, with a statistically significant increase of up to
25\% for 300~nm particles (Figure \ref{fig:sizedistrib}(a)). \jk{The large error bars in Figures~\ref{fig:mass1} and \ref{fig:sizedistrib}(c) are primarily due to fluctuations of the laser pulse energy (up to 20\%, amplified by the non-linear propagation), the fluctuations of the atmosphere around the laser during the measurements, as well as air currents in the wind-protecting tube (See Figure~\ref{fig:experimentalSetup}), which alter the efficiency with which laser-generated mass is sampled. Despite these, the results retain statistical significance. 
Note however that the error bars on the ratios are calculated as a worst-case scenario by adding the relative errors on each term, providing an upper limit to them.}

\subsection*{Size distribution}

\begin{figure}
\centerline{\includegraphics[width=0.7\columnwidth]{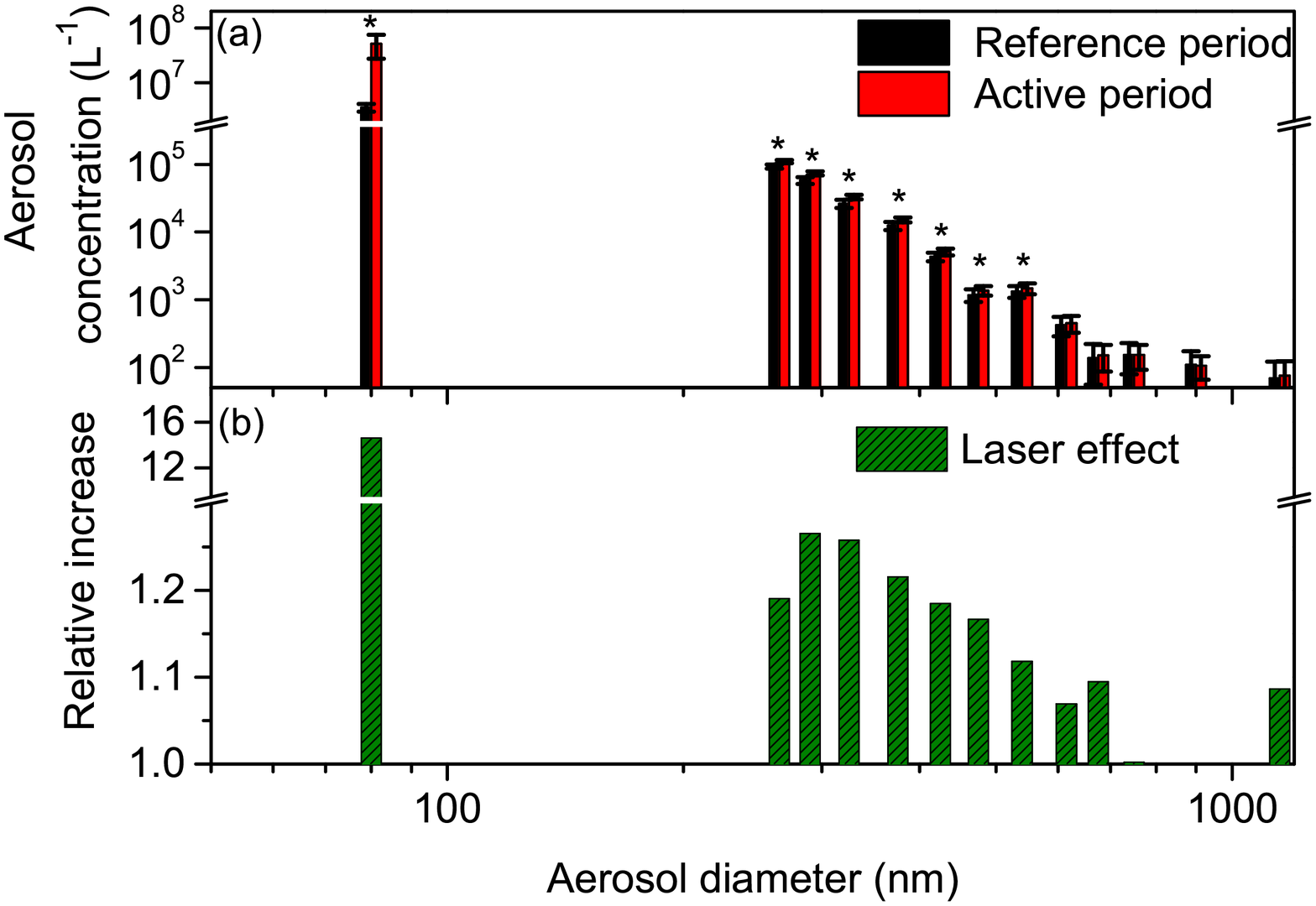}}
\centerline{\includegraphics[width=0.7\columnwidth]{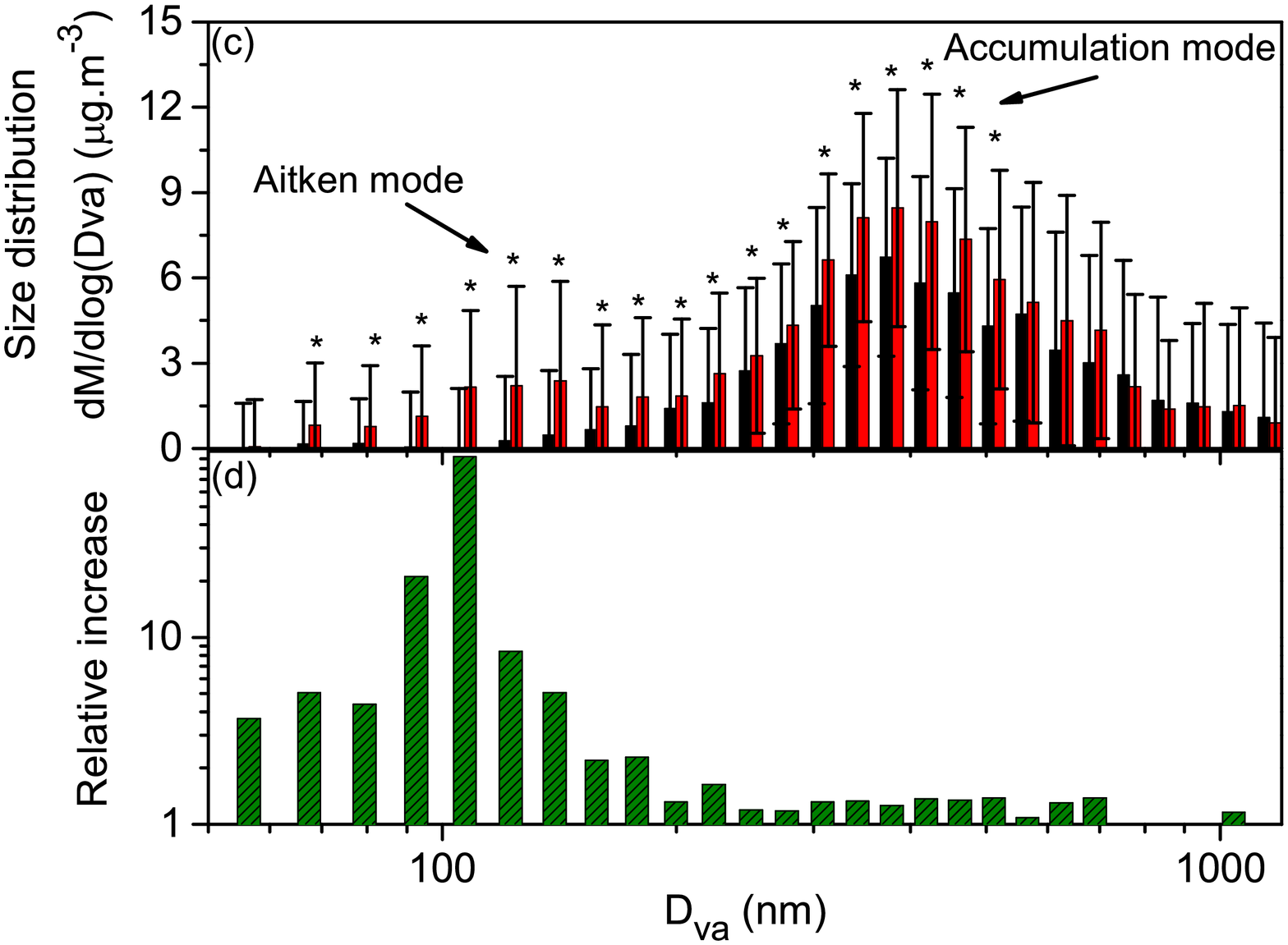}}
\caption{Evolution of the aerosol size distribution under laser illumination. (a) Aerosol size distribution between 250 and 900~nm during both reference
and active periods measured by the Grimm optical particle sizer, and concentration of 25~-~300~nm particles measured by the Nanocheck (column centered at the mean diameter 80~nm); (b) Corresponding relative increase induced by the laser; (c) Dry particle size distribution (total mass detected by the AMS, calculated as the sum of nitrate, sulfate, organics, ammonium and chloride) during both reference
and active periods
and (d) Corresponding relative increase induced by the laser. $D_\textrm{va}$ is the vacuum aerodynamic diameter. Stars denote size classes where the effect of the laser is statistically significant ($\alpha \le 0.01$, see text for details). \jk{Error bars correspond to one standard deviation.}
\label{fig:sizedistrib}}

\end{figure}

By looking at the total mass detected by the AMS (sum of the nitrate, sulfate, organics, ammonium and chloride), one can see the appearance of a new mode (Aitken mode) centered around 150~nm in the dry particle size distribution when the laser is turned on  (Figure~\ref{fig:sizedistrib}(c) and (d)). This mode could be formed either by nucleation or by growth of nanometric pre-existing particles. Simultaneously, the accumulation mode centered at 380~nm increases by 25\%.
also observed in the total mass size distribution.
In spite of the smaller size of the smaller mode, both modes contribute about  half of the total laser-induced mass condensation (Figure~\ref{fig:frac}(c)).

\subsection*{Chemical composition of laser-induced aerosols}

\begin{figure}
\centerline{\includegraphics[width=0.8\columnwidth]{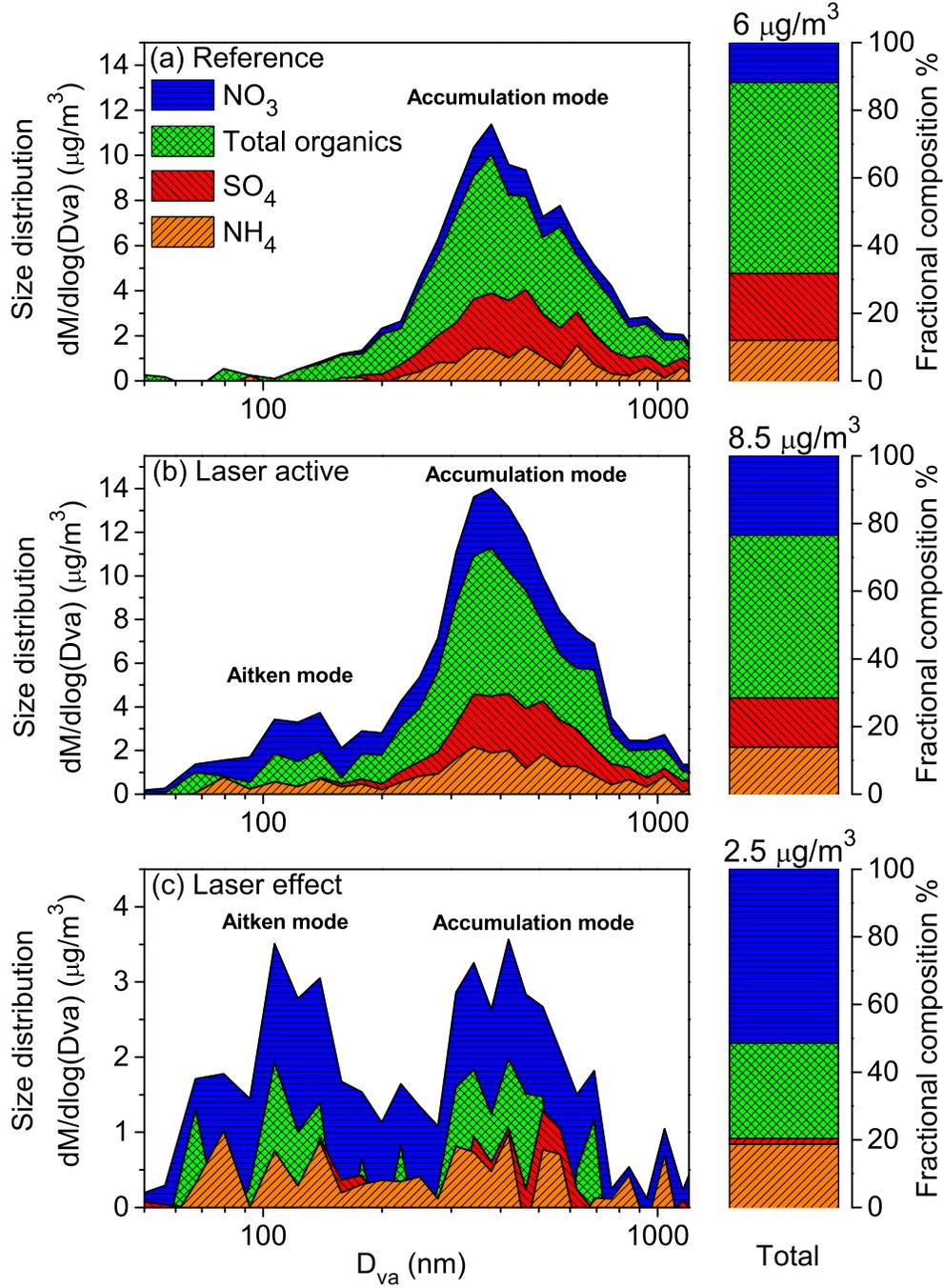}}
\caption{Mass distribution of the measured condensable species within the size distribution during (a) the reference period and (b) the laser active period. (c) Difference between the active and reference periods, displaying the condensation effect of the laser. $D_\textrm{va}$ is the vacuum aerodynamic diameter. 
\label{fig:frac} }
\end{figure}

Beyond the total particle mass and number concentration, the aerosol mass spectrometer allows identifying and quantifying the main components
that  substantially contribute to the laser-induced condensation.
The condensed mass of NO$_{3}^-$ increases by a factor of 2.8, NH$_{4}^+$ by 60\% , and organics by 20\% (Figure~\ref{fig:mass1}(b)). Considering their original mass fraction in the aerosols (Figure~\ref{fig:frac}a, right column),
these components contribute 50\%, 20\%, and 28\% of the net laser-induced
increase of the dry mass (Figure \ref{fig:frac}c, right column). These contributions to the laser-condensed mass are comparable in both size modes. 

In contrast, no evidence for significant laser-induced SO$_{4}^-$ condensation is observed, consistent with  the concentration of SO$_2$ in the atmosphere during the experiment of 2$\mathrm{\mu g / m^3} $ (as measured by the Geneva city air quality monitoring network \cite{transalpair}).

\subsection*{Oxygenation of the organic fraction of the aerosols}
\label{sub:oxidation-of-the}

The AMS allows detailed analysis of the organic aerosol composition. More specifically, oxygenated organic aerosol (OOA) can be described as highly oxygenated, low-volatility (LV-OOA), and less oxygenated semi-volatile OOA (SV-OOA)
\cite{ng_organic_2010,ng_changes_2011}. Considering the mass to charge ($m/z$) values
of 44 (corresponding in particular to $\mathrm{CO_{2}^{+}}$) and 43 (mainly
$\mathrm{C_{2}H_{3}O^{+}}$), as representative
of these two categories respectively, we investigated their relative contributions, hereafter denoted 
$f44$ and $f43$, to the total organic mass condensed
in the aerosols.

\begin{figure}
\centerline{\includegraphics[width=0.8\columnwidth]{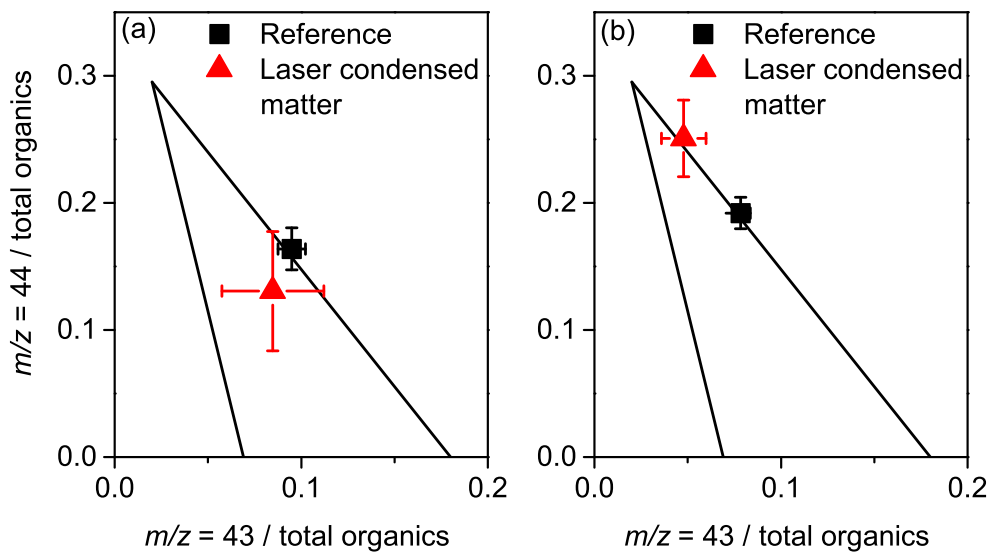} }
\caption{Fraction of highly ($m/z$~=~44, mainly $\mathrm{CO_{2}^{+}}$) and less oxygenated ($m/z$~=~43, mainly $\mathrm{C_{2}H_{3}O^{+}}$) organics among the total organics, in particles detected during the reference
period (black) and the laser-condensed particles, that is the difference of measurement with and without the laser (red), for (a) the main experiment and (b) when spraying water in a low RH atmosphere (b). The two
black lines encompass the region where most atmospheric data are found to date\cite{ng_organic_2010}.
\label{fig:44over43} }
\end{figure}

During the experiment, the laser condensed mass presents an $f44$ and $f43$ 25\% and 11\% respectively lower than in the pre-existing atmospheric aerosol (Figure \ref{fig:44over43}(a)). This oxygenation state of the laser-condensed organic mass is similar to typical ambient OOA\cite{ng_organic_2010}.

The laser produces highly oxidative conditions, with the generation
of local concentrations of $\mathrm{O_{3}}$ up to the 100~ppm range \cite{petit_production_2010,Camino2015} (see also Figure~\ref{fig:mass1})
and ppbs of $^\cdot$OH radicals  in the filaments \cite{matthews_cooperative_2013}. 
These conditions produce OOA with a composition on the boundary of LV-OOA and SV-OOA from the gaseous organics within a few seconds.

\subsection*{Spraying water droplets to a low RH atmosphere}

In an atmosphere with 50\% RH, the laser doesn't condense any mass, and the ambient aerosol is mainly composed of organics (68\%), SO$_4$ (19\%), NH$_4$ (8\%) and a small percentage of NO$_3$ (5\%). Continuously spraying water particles close to the laser focus caused the laser to condense mass (as evidenced by a mass increase of 8\%), mainly composed of organics (85\%). Furthermore, the oxygenation of the laser-condensed organics is higher than the pre-existing ones in this case, as evidenced by the increase of the highly oxygenated organics (see $f44$ in Figure \ref{fig:44over43}(b)) while the less oxygenated organics ($f43$) simultaneously decrease. The huge amount of $^{\bullet}$OH produced by the interaction of the water droplets with the laser filaments results in the production of LV-OOA, much more oxygenated than in the main experiment. Conversely, no condensation of hygroscopic species occurs despite the huge amount of water sprayed, indicating that the latter is not available for laser-induced condensation.

\subsection*{Influence of laser pulse energy}

\jk{We investigated the effect of the laser pulse energy on the efficiency of laser-induced condensation.
Reducing the laser pulse energy from 180~mJ to 110~mJ did
not visibly alter the effect of the laser on ambient aerosol.
In contrast, at 55~mJ per pulse no effect on the ambient aerosol (i.e. no production of aerosol mass) was observed.}

\section*{Discussion}
\subsection*{Laser-induced condensation}

The formation of a new Aitken mode accounting for 50\% of the laser-condensed dry mass together with the increase of the pre-existing mode provides a clear evidence of the condensation induced by the laser filaments in the atmosphere, an effect that had only been characterised before in laboratory conditions~\cite{saathoff_laser_2013}. 
The formation of new particles by nucleation or growth of pre-existing nanometric particles, as opposed to shattering of larger particles, is confirmed by the increase of the total mass of aerosol detected during the experiment, by the change of chemical composition of their dry mass and by the lower oxygenation level of the organics contained in the particles detected under laser illumination.

\subsection*{Condensation of ammonium nitrate}
The concentration of NO$_{3}$ and NH$_4$ observed in the laser-induced aerosols (50\% and almost 20\% of the dry condensed mass, respectively, i.e., a total contribution close to 70\%) sheds new light on the  previously inferred binary water-HNO$_{3}$  condensation under laser illumination.
The 0.31 mass ratio of $\mathrm{NH_{4}}$ to $\mathrm{NO_{3}}$ in the dry condensed mass (Figure \ref{fig:frac}), in both the nucleation mode and accumulation modes, is consistent with that of $\mathrm{NH_{4}NO_{3}}$.
One can therefore expect that the hygroscopic HNO$_3$ previously identified to assist the condensation of water \cite{henin_field_2011,rohwetter_modelling_2011} is not condensed, as initially expected, as nitric acid in a binary mixture with water, but rather as hygroscopic ammonium nitrate through the reaction \cite{seinfeld_atmospheric_2006}

\begin{equation}
\mathrm{NH_{3}+HNO_{3}\rightleftharpoons NH_{4}NO_{3}}
\label{eq:NH4NO3}
\end{equation}

NH$_3$ is typically available as background trace gas in the summer sub-urban atmosphere of the experiment location in amounts of several $\mu$g/m$^3$~\cite{Thoni2004}. This concentration is comparable with the laser-condensed mass of NH$_4^+$, suggesting that the laser-induced condensation relies on the condensation of pre-existing ammonia together with HNO$_{3}$. The latter is generated via the interaction of NO$_{x}$ produced by the multi-photon ionisation of N$_2$ with O$_3$ \cite{petit_production_2010} or $^\cdot$OH radicals \cite{matthews_cooperative_2013}, both produced by the multiphoton absorption and photolysis of O$_2$.
The very low efflorescence of ammonium nitrate and the high quantities of $\mathrm{NH_{4}NO_{3}}$ in the laser condensed mass suggest that there is always water in the laser induced aerosols.
Besides, it validates a posteriori the \emph{ad hoc} introduction of ammonium nitrate for modelling the growth of laser-induced particles \cite{rohwetter_modelling_2011}.

\subsection*{Organics}
Besides ammonium nitrate, the laser mainly condenses organics (28\% of the dry mass). 
The amount of organics condensed by the laser seems to be driven by the ability of the laser to create an oxidative atmosphere. At normal atmospheric conditions and when RH is above 70\%, the laser partly oxidises organics, resulting in the condensation of typical ambient LV/SV-OOA, slightly less oxygenated than the pre-existing organic aerosol during our experiment. This suggests that new aerosol mass
is formed by the laser from the semi-volatile organics available in
the gas phase, and not from the already condensed non-volatile species.
The absolute amount slightly below 1~$\mu$g/m$^3$ of organics condensed by the laser constitutes a few percent of the total concentration of VOCs available in a typical urban atmosphere \cite{Derwent2014}, confirming the VOC availability for laser-induced oxidation and condensation.

When water is sprayed around the laser filaments at an original RH lower than 70\%, the proportion of organics in the dry mass drastically increases (80\% of the dry mass). Given that no relative humidity change was observed when spraying water during this particular experiment, we attribute this effect to the interaction of the water droplets sprayed into the laser beam, with the laser filaments. Considering their high concentration during the experiment, about 100 water droplets interact  with the laser filaments for each laser pulse. These water droplets then internally focus the incident laser light on a nanometric hot spot, where the extreme incident intensity (in the 10$^{15}$~TW/cm$^2$ range) will efficiently ionize water \cite{hill_enhanced_2000}. This ionization produces  a concentration of  $^\cdot$OH radicals much higher than in normal atmospheric conditions, leading to efficient oxidization of the available organics. The resulting highly oxygenated, low volatility organics then efficiently condense onto  particles with  a high oxygenation state comparable to that of $\alpha$-pinene particles after exposure to 500 ppb
of O$_{3}$ during 13 hours \cite{shilling_particle_2008,ng_organic_2010}. This very fast oxidization illustrates again the extreme conditions locally produced by the laser filaments.

Our results evidence the dual role of organics in
laser-induced condensation. While they had been shown to promote nucleation
in a background-free atmosphere \cite{saathoff_laser_2013}, we show
that they also substantially contribute to the particle growth up
to at least 600~nm in diameter (see Figure \ref{fig:frac}).

\subsection*{Sulfate}

Sulfuric acid is known to provide one of the main sources of condensation nuclei in the atmsophere \cite{seinfeld_atmospheric_2006}.
In the case of laser-induced nucleation, experiments in a cloud chamber showed that sulfuric acid does not only drastically enhance the number density of nucleated particles, but also the total condensed mass \cite{saathoff_laser_2013}. 
The very low mass fraction of sulfate ions in the laser-induced condensed mass
($<$ 2 \%) detected during our experiment can be explained by the low ambient SO$_2$ concentrations during the study.
In this case, its role is restricted to the formation of small condensation nuclei, most probably together with oxidized organics \cite{riccobono_oxidation_2014}.

\subsection*{Laser energy}
\jk{The energy dependence of the efficiency of laser-induced condensation appears to display a threshold between 50 and 110~mJ. This suggests that the effect of the laser can be understood by considering the interplay between two phenomena. }

\jk{On one side, the efficiency of the plasma photochemistry increases non-linearly with increasing incident laser pulse intensity. At 800 nm, the photolysis of O$_2$, its ionization, and that of N$_2$ respectively require 4, 8, and 11 photons \cite{couairon_femtosecond_2007}.
The associated efficiencies therefore depend on the corresponding powers of the laser input intensity, so that the condensation mechanisms initiated by these processes will successively come into play when the laser energy rises. This results in the observed dependence with the pulse energy.
On the other side, however, particle formation and growth requires the availability of condensable
species in the volume swept by the high intensity laser, or at reach within a reasonable transport time.
As this transport mainly occurs by diffusion,
which is very slow for distances beyond the mm-range, diffusion constitutes the limiting
 factor to condensation when the laser pulse energy is further increased.}

\jk{This finding illustrates that increasing the magnitude of the laser
effect requires to increase the activated volume, by either increasing
the beam diameter, or by steering the beam dynamically.}

\section*{Conclusion}

We simultaneously measured  the size distribution and the quantitative size resolved aerosol composition by mass spectrometry during laser-induced condensation in the real atmosphere. We obtained direct evidence of laser-assisted condensation in real atmosphere, with a 40\% increase of aerosol total mass and a 15-fold increase in the particle number concentration between 25 and 300~nm. Furthermore, we observed the condensation of hygroscopic NH$_4$NO$_3$ accounting for two thirds of the laser-condensed mass. This abundance shows that NO$_{3}$ condenses as ammonium nitrate, rather than as nitric acid as previously expected. The rest of the laser-condensed dry mass is mainly provided by organics, accounting for as much as 28\%, evidencing their role in the particle growth.
The presence of these two components (ammonium nitrate and organics) can be attributed to two main condensation mechanisms related to the production of  nitrogen oxides, ozone and $^\cdot$OH radicals by the interaction of the filaments with the atmosphere. 
The first one consists of the generation of HNO$_3$, which then condenses in a ternary system together with NH$_3$ and water to form 
$\mathrm{NH_4NO_3}$ in the laser-induced particles. The second one is the oxidation of volatile organics  by $^\cdot$OH radicals and O$_3$ into less volatile organics which condense on the pre-existing particles. 
Moreover, the oxidation as well as the degree of oxygenation of organics can be strongly increased by the presence of pre-existing water droplets because of increased $^\cdot$OH radical concentrations due to internal focusing of the laser light within the droplets. Our work therefore offers a global picture of the complex mechanisms at play in laser-induced condensation. 

\section*{Methods}

\paragraph*{Teramobile laser:}
The Teramobile \cite{wille_teramobile:_2002} is a Ti:sapphire laser delivering 80~fs pulses of 180~mJ, at a repetition rate of 10 Hz with a wavelength centered at 800~nm. The beam is sent horizontally
into the open atmosphere, 120~cm above ground. The beam has an initial
diameter of 7.5~cm and is loosely focused by an $f=25$~m telescope.
The beam at the focus has a waist of around 5~mm (HWHM) and contains
$\sim$ 30 filaments. The beam is blocked by a screen after a further
15~m of propagation to avoid any interference by particles ablated
from this beam dump.
\paragraph*{Wind-protecting tube:}
The region around the focus is shielded from  wind by a 2 m long,
30 cm diameter galvanized iron tube. It is
open at each end, with 10~cm diameter ports letting both the laser
beam get through and exchanges with the free atmosphere to occur.
The detection instruments continuously monitored the atmosphere at 1~cm distance
from the laser beam close to the filaments, via 1.5 m-long, 4~mm inner diameter
copper sampling tubes connected to the center of this shielding tube.
\paragraph*{AMS:}
Size-resolved non-refractory particle composition was measured using an Aerodyne high-resolution time-of-flight aerosol mass spectrometer \jk{(AMS)}~\cite{decarlo_particle_2004,canagaratna_chemical_2007}. Briefly, the AMS continuously samples particles from ambient air (0.8 L/min) through a 100 $\mathrm{\mu m}$ critical orifice into an aerodynamic lens (2 torr). Here the particles are focused into a narrow beam and accelerated to a velocity that is inversely related to their vacuum aerodynamic diameter. The particle beam can be either alternately blocked and unblocked (yielding ensemble mass spectra of the incident aerosol), or modulated by a spinning chopper wheel (150 Hz), yielding size-resolved mass spectra at the cost of lower sensitivity. The particle ionisation on a resistively heated surface (600\textdegree C, $\mathrm{10^{-7}}$ torr), where the non-refractory components flash vaporize, are ionized by electron impact (70 eV) and are detected by time-of-flight mass spectrometry. The AMS can detect most atmospherically-relevant species except for black carbon and mineral dust, while water is not quantitatively detected due to high instrument backgrounds and uncertainty over the fraction of water that evaporates in the aerodynamic lens. The lens has unit transmission efficiency for particles with vacuum aerodynamic diameters between 100 and 600 nm, and approximately 10\% transmission of 50 and 1000 nm particles \cite{liu_transmission_2007}.  Mass spectra are analyzed and quantitatively split into e.g. nitrate, sulfate, ammonium, and organics using the standard fragmentation table-based analysis of Allan et al. \cite{allan_generalised_2004}. Depending on the experiment, the AMS was operated with either 1 min or 20 sec time resolution.
\paragraph*{Particle sizer:}
The aerosol sizer is a Grimm 1.109 measuring the aerosol size distribution by optical scattering in 31 classes ranging from 250~nm to 30~$\mu$m. 
This instrument samples 1.5~L/min
and provides measurement at a temporal resolution of 6~s, although
the overall reaction time of the instrument and the associated sampling
tube can be estimated to lie in the 8~s range. It also integrates
the size distribution up to 1 and 2.5~$\mu$m diameter, respectively, to estimate the PM1 and PM2.5 mass concentration in air. 
\paragraph*{Nanoparticle sensor}
A nanoparticle sensor (Grimm Nanocheck 1.365) measures
the 25--300~nm particle number density. Sampling flow and temporal resolution are the same as the above particle sizer.
\paragraph*{Ozone detector:}
The ozone detector is a Horiba APOA-350E and sampled the atmosphere \jk{at a rate of} 2~L/min, measuring ozone concentration with a temporal resolution of 20~s. 
\paragraph*{Temperature and RH probe:}
The device monitoring temperature and RH is a Testo 635-2 device. Its probe is located at the same place as the other instruments.
\paragraph*{Nebuliser:}
The nebuliser used to spray water is a TSI 3076 nebuliser equipped with a 3074B air filter. The corresponding size distribution was centred around 1.5 $\mathrm{\mu m}$, with a geometric standard deviation smaller than 2.

\section*{Acknowledgements}

D. Mongin acknowledges S. Hermelin for his permanent help and fruitful discussions.
This research was supported by the ERC advanced grant "Filatmo", the FP7 ITN network "JMAP" and the SNSF starting grant "BSSGI0 155846".

\section*{Author Contributions}
JPW, US, AS, and JK designed the experiment; DM, JS, ES, and MM designed and set up the experimental setup; DM, JS, ES, JGB, NB, and JK carried out the experiments; DM, JS, and JK designed and carried out the data analysis; DM, JS, and JK wrote the paper; All authors reviewed the manuscript.

\section*{Competing financial interests}
The authors declare no competing financial interests.


\begin{thebibliography}{10}
\expandafter\ifx\csname url\endcsname\relax
  \def\url#1{\texttt{#1}}\fi
\expandafter\ifx\csname urlprefix\endcsname\relax\def\urlprefix{URL }\fi
\providecommand{\bibinfo}[2]{#2}
\providecommand{\eprint}[2][]{\url{#2}}

\bibitem{kelley_self-focusing_1965}
\bibinfo{author}{Kelley, P.}
\newblock \bibinfo{title}{Self-focusing of optical beams}.
\newblock \emph{\bibinfo{journal}{Phys. Rev. Lett.}}
  \textbf{\bibinfo{volume}{15}}, \bibinfo{pages}{1005--1008}
  (\bibinfo{year}{1965}).

\bibitem{braun_self-channeling_1995}
\bibinfo{author}{Braun, A.} \emph{et~al.}
\newblock \bibinfo{title}{Self-channeling of high-peak-power femtosecond laser
  pulses in air}.
\newblock \emph{\bibinfo{journal}{Opt. Lett.}}
  \textbf{\bibinfo{volume}{20}}, \bibinfo{pages}{73--75}
  (\bibinfo{year}{1995}).

\bibitem{couairon_femtosecond_2007}
\bibinfo{author}{Couairon, A.} \& \bibinfo{author}{Mysyrowicz, A.}
\newblock \bibinfo{title}{Femtosecond filamentation in transparent media}.
\newblock \emph{\bibinfo{journal}{Phys. Rep.}}
  \textbf{\bibinfo{volume}{441}}, \bibinfo{pages}{47--189}
  (\bibinfo{year}{2007}).

\bibitem{berge_ultrashort_2007}
\bibinfo{author}{Berg{\'e}, L.}, \bibinfo{author}{Skupin, S.},
  \bibinfo{author}{Nuter, R.}, \bibinfo{author}{Kasparian, J.} \&
  \bibinfo{author}{Wolf, J.}
\newblock \bibinfo{title}{Ultrashort filaments of light in weakly-ionized,
  optically-transparent media}.
\newblock \emph{\bibinfo{journal}{Rep. Prog. Phys.}}
  \textbf{\bibinfo{volume}{70}}, \bibinfo{pages}{1633--1713}
  (\bibinfo{year}{2007}).

\bibitem{chin_propagation_2005}
\bibinfo{author}{Chin, S.~L.} \emph{et~al.}
\newblock \bibinfo{title}{The propagation of powerful femtosecond laser pulses
  in optical media: physics, applications, and new challenges}.
\newblock \emph{\bibinfo{journal}{Can. J. Phys.}}
  \textbf{\bibinfo{volume}{83}}, \bibinfo{pages}{863--905}
  (\bibinfo{year}{2005}).

\bibitem{bejot_higher-order_2010}
\bibinfo{author}{B\'ejot, P.} \emph{et~al.}
\newblock \bibinfo{title}{Higher-order kerr terms allow ionization-free
  filamentation in gases}.
\newblock \emph{\bibinfo{journal}{Phys. Rev. Lett.}}
  \textbf{\bibinfo{volume}{104}}, \bibinfo{pages}{103903}
  (\bibinfo{year}{2010}).

\bibitem{bejot_transition_2011}
\bibinfo{author}{B{\'e}jot, P.} \emph{et~al.}
\newblock \bibinfo{title}{Transition from plasma- to kerr-driven laser
  filamentation}.
\newblock \emph{\bibinfo{journal}{Phys. Rev. Lett.}}
  \textbf{\bibinfo{volume}{106}}, \bibinfo{pages}{243902}
  (\bibinfo{year}{2011}).

\bibitem{la_fontaine_filamentation_1999}
\bibinfo{author}{La~Fontaine, B.} \emph{et~al.}
\newblock \bibinfo{title}{Filamentation of ultrashort pulse laser beams
  resulting from their propagation over long distances in air}.
\newblock \emph{\bibinfo{journal}{Phys. Plasmas}}
  \textbf{\bibinfo{volume}{6}}, \bibinfo{pages}{1615--1621}
  (\bibinfo{year}{1999}).

\bibitem{rodriguez_kilometer-range_2004}
\bibinfo{author}{Rodriguez, M.} \emph{et~al.}
\newblock \bibinfo{title}{Kilometer-range nonlinear propagation of femtosecond
  laser pulses}.
\newblock \emph{\bibinfo{journal}{Phys. Rev.  E}}
  \textbf{\bibinfo{volume}{69}}, \bibinfo{pages}{036607}
  (\bibinfo{year}{2004}).
  
\bibitem{mejean_multifilamentation_2005}
\bibinfo{author}{M\'ejean, G.} \emph{et~al.}
\newblock \bibinfo{title}{Multifilamentation transmission through fog}.
\newblock \emph{\bibinfo{journal}{Phys. Rev.  E}}
  \textbf{\bibinfo{volume}{72}}, \bibinfo{pages}{026611}
  (\bibinfo{year}{2005}).

\bibitem{chin_filamentation_2002}
\bibinfo{author}{Chin, S.~L.} \emph{et~al.}
\newblock \bibinfo{title}{Filamentation of femtosecond laser pulses in
  turbulent air}.
\newblock \emph{\bibinfo{journal}{Appl. Phys. B}}
  \textbf{\bibinfo{volume}{74}}, \bibinfo{pages}{67--76}
  (\bibinfo{year}{2002}).

\bibitem{salame_propagation_2007}
\bibinfo{author}{Salam{\'e}, R.}, \bibinfo{author}{Lascoux, N.},
  \bibinfo{author}{Salmon, E.}, \bibinfo{author}{Kasparian, J.} \&
  \bibinfo{author}{Wolf, J.}
\newblock \bibinfo{title}{Propagation of laser filaments through an extended
  turbulent region}.
\newblock \emph{\bibinfo{journal}{Appl. Phys. Lett.}}
  \textbf{\bibinfo{volume}{91}}, \bibinfo{pages}{171106--171106}
  (\bibinfo{year}{2007}).

\bibitem{kasparian_physics_2008}
\bibinfo{author}{Kasparian, J.} \& \bibinfo{author}{Wolf, J.-P.}
\newblock \bibinfo{title}{Physics and applications of atmospheric nonlinear
  optics and filamentation}.
\newblock \emph{\bibinfo{journal}{Opt. Express}}
  \textbf{\bibinfo{volume}{16}}, \bibinfo{pages}{466--493}
  (\bibinfo{year}{2008}).

\bibitem{kasparian_white-light_2003}
\bibinfo{author}{Kasparian, J.} \emph{et~al.}
\newblock \bibinfo{title}{White-light filaments for atmospheric analysis}.
\newblock \emph{\bibinfo{journal}{Science}} \textbf{\bibinfo{volume}{301}},
  \bibinfo{pages}{61--64} (\bibinfo{year}{2003}).

\bibitem{Dogariu2011}
\jk{Dogariu, A., Michael, J. B., Scully,  M. O., Miles, R. B. High-Gain Backward Lasing in Air, \textit{Science} \textbf{331}, 442--445 (2011)}

\bibitem{Yao2011}
Yao, J. et al. 
High-brightness switchable multiwavelength remote laser in air, \textit{Phys. Rev. A} \textbf{84}, 051802 (2011)

\bibitem{zhao_femtosecond_1995}
\bibinfo{author}{Zhao, X.}, \bibinfo{author}{Diels, J.}, \bibinfo{author}{Wang,
  C.} \& \bibinfo{author}{Elizondo, J.}
\newblock \bibinfo{title}{Femtosecond ultraviolet laser pulse induced lightning
  discharges in gases}.
\newblock \emph{\bibinfo{journal}{{IEEE} J. Quantum Electron.}}
  \textbf{\bibinfo{volume}{31}}, \bibinfo{pages}{599--612}
  (\bibinfo{year}{1995}).

\bibitem{pepin_triggering_2001}
\bibinfo{author}{P\'epin, H.} \emph{et~al.}
\newblock \bibinfo{title}{Triggering and guiding high-voltage large-scale
  leader discharges with sub-joule ultrashort laser pulses}.
\newblock \emph{\bibinfo{journal}{Phys. Plasmas}}
  \textbf{\bibinfo{volume}{8}}, \bibinfo{pages}{2532--2539}
  (\bibinfo{year}{2001}).

\bibitem{rodriguez_megavolt_2002}
\bibinfo{author}{Kasparian, J.} \emph{et~al.}
\newblock \bibinfo{title}{Megavolt discharges triggered and guided with laser
  filaments}.
\newblock \emph{\bibinfo{journal}{Opt. Lett.}}
  \textbf{\bibinfo{volume}{27}}, \bibinfo{pages}{772--774}
  (\bibinfo{year}{2002}).

\bibitem{kasparian_electric_2008}
\bibinfo{author}{Kasparian, J.} \emph{et~al.}
\newblock \bibinfo{title}{Electric events synchronized with laser filaments in
  thunderclouds}.
\newblock \emph{\bibinfo{journal}{Opt. Express}}
  \textbf{\bibinfo{volume}{16}}, \bibinfo{pages}{5757--5763}
  (\bibinfo{year}{2008}).

\bibitem{rohwetter_laser-induced_2010}
\bibinfo{author}{Rohwetter, P.} \emph{et~al.}
\newblock \bibinfo{title}{Laser-induced water condensation in air}.
\newblock \emph{\bibinfo{journal}{Nature Photon.}}
  \textbf{\bibinfo{volume}{4}}, \bibinfo{pages}{451--456}
  (\bibinfo{year}{2010}).

\bibitem{henin_field_2011}
\bibinfo{author}{Henin, S.} \emph{et~al.}
\newblock \bibinfo{title}{Field measurements suggest the mechanism of
  laser-assisted water condensation}.
\newblock \emph{\bibinfo{journal}{Nature Comm.}}
  \textbf{\bibinfo{volume}{2}}, \bibinfo{pages}{456} (\bibinfo{year}{2011}).

\bibitem{kasparian_laser-assisted_2012}
\bibinfo{author}{Kasparian, J.}, \bibinfo{author}{Rohwetter, P.},
  \bibinfo{author}{W{\"o}ste, L.} \& \bibinfo{author}{Wolf, J.}
\newblock \bibinfo{title}{Laser-assisted water condensation in the atmosphere:
  a step towards modulating precipitation?}
\newblock \emph{\bibinfo{journal}{J. Phys. D: Appl. Phys.}}
  \textbf{\bibinfo{volume}{45}}, \bibinfo{pages}{293001}
  (\bibinfo{year}{2012}).

\bibitem{leisner_laser-induced_2013}
\bibinfo{author}{Leisner, T.} \emph{et~al.}
\newblock \bibinfo{title}{Laser-induced plasma cloud interaction and ice
  multiplication under cirrus cloud conditions}.
\newblock \emph{\bibinfo{journal}{Proc. Nat. Acad.
  Sci.}} \textbf{\bibinfo{volume}{110}}, \bibinfo{pages}{10106--10110}
  (\bibinfo{year}{2013}).

\bibitem{SunLWJWWGLCLX2013}
\bibinfo{author}{Sun, H.} \emph{et~al.}
\newblock \bibinfo{title}{Laser filamentation induced air-flow motion in a
  diffusion cloud chamber}.
\newblock \emph{\bibinfo{journal}{Opt. Express}}
  \textbf{\bibinfo{volume}{21}}, \bibinfo{pages}{9255--9266}
  (\bibinfo{year}{2013}).


\bibitem{Ju2014}
\bibinfo{author}{Ju, J.} \emph{et~al.}
\newblock \bibinfo{title}{Laser-induced supersaturation and snow formation in a
  sub-saturated cloud chamber}.
\newblock \emph{\bibinfo{journal}{Appl. Phys. B}}
  \textbf{\bibinfo{volume}{117}}, \bibinfo{pages}{1001--1007}
  (\bibinfo{year}{2014}).


\bibitem{saathoff_laser_2013}
\bibinfo{author}{Saathoff, H.} \emph{et~al.}
\newblock \bibinfo{title}{Laser filament-induced aerosol formation}.
\newblock \emph{\bibinfo{journal}{Atmos. Chem. Phys.}}
  \textbf{\bibinfo{volume}{13}}, \bibinfo{pages}{4593--4604}
  (\bibinfo{year}{2013}).


\bibitem{rohwetter_modelling_2011}
\bibinfo{author}{Rohwetter, P.}, \bibinfo{author}{Kasparian, J.},
  \bibinfo{author}{W{\"o}ste, L.} \& \bibinfo{author}{Wolf, J.}
\newblock \bibinfo{title}{Modelling of {HNO}$_3$-mediated laser-induced
  condensation: A parametric study}.
\newblock \emph{\bibinfo{journal}{J. Chem. Phys.}}
  \textbf{\bibinfo{volume}{135}}, \bibinfo{pages}{134703}
  (\bibinfo{year}{2011}).


\bibitem{petit_production_2010}
\bibinfo{author}{Petit, Y.}, \bibinfo{author}{Henin, S.},
  \bibinfo{author}{Kasparian, J.} \& \bibinfo{author}{Wolf, J.}
\newblock \bibinfo{title}{Production of ozone and nitrogen oxides by laser
  filamentation}.
\newblock \emph{\bibinfo{journal}{Appl. Phys. Lett.}}
  \textbf{\bibinfo{volume}{97}}, \bibinfo{pages}{021108}
  (\bibinfo{year}{2010}).
  

\bibitem{wille_teramobile:_2002}
\bibinfo{author}{Wolf, J.-P.} \emph{et~al.}
\newblock \bibinfo{title}{Teramobile: a mobile femtosecond-terawatt laser and
  detection system}.
\newblock \emph{\bibinfo{journal}{Eur. Phys. J. - Appl. Phys.}}
  \textbf{\bibinfo{volume}{20}}, \bibinfo{pages}{183--190}
  (\bibinfo{year}{2002}).

\bibitem{transalpair}
\bibinfo{author}{Transalp'AIR}. Indice de qualité de l'air.
\newblock \urlprefix\url{www.transalpair.eu}. Date of access 24/08/2015.

\bibitem{ng_organic_2010}
\bibinfo{author}{Ng, N.}, \bibinfo{author}{Canagaratna, M.},
  \bibinfo{author}{Zhang, Q.}, \bibinfo{author}{Jimenez, J.} \&
  \bibinfo{author}{Tian, J. E.~A.}
\newblock \bibinfo{title}{Organic aerosol components observed in northern
  hemispheric datasets from aerosol mass spectrometry}.
\newblock \emph{\bibinfo{journal}{Atmos. Chem. Phys.}}
  \textbf{\bibinfo{volume}{10}}, \bibinfo{pages}{4625--4641}
  (\bibinfo{year}{2010}).


\bibitem{ng_changes_2011}
\bibinfo{author}{Ng, N.} \emph{et~al.}
\newblock \bibinfo{title}{Changes in organic aerosol composition with aging
  inferred from aerosol mass spectra}.
\newblock \emph{\bibinfo{journal}{Atmos. Chem. Phys.}}
  \textbf{\bibinfo{volume}{11}}, \bibinfo{pages}{6465--6474}
  (\bibinfo{year}{2011}).


\bibitem{Camino2015}
\bibinfo{author}{Camino, A.}, \bibinfo{author}{Li, S.}, \bibinfo{author}{Hao,
  Z.} \& \bibinfo{author}{Lin, J.}
\newblock \bibinfo{title}{Spectroscopic determination of NO$_2$, NO$_3$, and O$_3$
  temporal evolution induced by femtosecond filamentation in air}.
\newblock \emph{\bibinfo{journal}{Appl. Phys. Lett.}}
  \textbf{\bibinfo{volume}{106}}, \bibinfo{pages}{021105}
  (\bibinfo{year}{2015}).


\bibitem{matthews_cooperative_2013}
\bibinfo{author}{Matthews, M.} \emph{et~al.}
\newblock \bibinfo{title}{Cooperative effect of ultraviolet and near-infrared
  beams in laser-induced condensation}.
\newblock \emph{\bibinfo{journal}{Appl. Phys. Lett.}}
  \textbf{\bibinfo{volume}{103}}, \bibinfo{pages}{264103}
  (\bibinfo{year}{2013}).


\bibitem{seinfeld_atmospheric_2006}
\bibinfo{author}{Seinfeld, J.} \& \bibinfo{author}{Pandis, S.}
\newblock \emph{\bibinfo{title}{Atmospheric Chemistry and
  Physics{\textemdash}From Air Pollution to Climate Change}}
  (\bibinfo{publisher}{Hoboken}, \bibinfo{year}{2006}), \bibinfo{edition}{2nd}
  edition.

\bibitem{Thoni2004}
\bibinfo{author}{Th\"oni, L.}, \bibinfo{author}{Brang, P.},
  \bibinfo{author}{Braun, S.}, \bibinfo{author}{Seitler, E.} \&
  \bibinfo{author}{Rihm, B.}
\newblock \bibinfo{title}{Ammonia monitoring in {Switzerland} with passive
  samplers: patterns, determinants and comparison with modelled
  concentrations}.
\newblock \emph{\bibinfo{journal}{Env. Monitoring Assess.}}
  \textbf{\bibinfo{volume}{98}}, \bibinfo{pages}{93--107}
  (\bibinfo{year}{2004}).

\bibitem{Derwent2014}
\bibinfo{author}{Derwent, R.~G.} \emph{et~al.}
\newblock \bibinfo{title}{Twenty years of continuous high time resolution
  volatile organic compound monitoring in the {United} {Kingdom} from 1993 to
  2012}.
\newblock \emph{\bibinfo{journal}{Atmos. Env.}}
  \textbf{\bibinfo{volume}{99}}, \bibinfo{pages}{239--247}
  (\bibinfo{year}{2014}).


\bibitem{hill_enhanced_2000}
\bibinfo{author}{Hill, S.~C.} \emph{et~al.}
\newblock \bibinfo{title}{Enhanced backward-directed multiphoton-excited
  fluorescence from dielectric microspheres}.
\newblock \emph{\bibinfo{journal}{Phys. Rev. Lett.}}
  \textbf{\bibinfo{volume}{85}}, \bibinfo{pages}{54--57}
  (\bibinfo{year}{2000}).

\bibitem{shilling_particle_2008}
\bibinfo{author}{Shilling, J.} \emph{et~al.}
\newblock \bibinfo{title}{Particle mass yield in secondary organic aerosol
  formed by the dark ozonolysis of $\alpha$-pinene}.
\newblock \emph{\bibinfo{journal}{Atmos. Chem. Phys.}}
  \textbf{\bibinfo{volume}{8}}, \bibinfo{pages}{2073--2088}
  (\bibinfo{year}{2008}).


\bibitem{kulmala_stable_2000}
\bibinfo{author}{Kulmala, M.}, \bibinfo{author}{Pirjola, L.} \&
  \bibinfo{author}{M{\"a}kel{\"a}, J.}
\newblock \bibinfo{title}{Stable sulphate clusters as a source of new
  atmospheric particles}.
\newblock \emph{\bibinfo{journal}{Nature}} \textbf{\bibinfo{volume}{404}},
  \bibinfo{pages}{66--69} (\bibinfo{year}{2000}).


\bibitem{decarlo_particle_2004}
\jk{DeCarlo, P., Kimmel, J., Trimborn, A., Northway, M., Jayne, J., Aiken, A., Gonin, M., Fuhrer, K., Horvath, T., Docherty, K., Worsnop, D., and Jimenez, J. Field-deployable, high-resolution, time-of-flight aerosol mass spectrometer, \textit{Anal. Chem.}, \textbf{78}, 8281-8289 (2006).}


\bibitem{canagaratna_chemical_2007}
\bibinfo{author}{Canagaratna, M.} \emph{et~al.}
\newblock \bibinfo{title}{Chemical and microphysical characterization of
  ambient aerosols with the Aerodyne aerosol mass spectrometer}.
\newblock \emph{\bibinfo{journal}{Mass Spec. Rev.}}
  \textbf{\bibinfo{volume}{26}}, \bibinfo{pages}{185--222}
  (\bibinfo{year}{2007}).

  

\bibitem{liu_transmission_2007}
\bibinfo{author}{Liu, P.} \emph{et~al.}
\newblock \bibinfo{title}{Transmission efficiency of an aerodynamic focusing
  lens system: Comparison of model calculations and laboratory measurements for
  the aerodyne aerosol mass spectrometer}.
\newblock \emph{\bibinfo{journal}{Aerosol Sci. Technol.}}
  \textbf{\bibinfo{volume}{41}}, \bibinfo{pages}{721--733}
  (\bibinfo{year}{2007}).


\bibitem{allan_generalised_2004}
\bibinfo{author}{Allan, J.} \emph{et~al.}
\newblock \bibinfo{title}{A generalised method for the extraction of chemically
  resolved mass spectra from Aerodyne aerosol mass spectrometer data}.
\newblock \emph{\bibinfo{journal}{J. Aerosol Sci.}}
  \textbf{\bibinfo{volume}{35}}, \bibinfo{pages}{909--922}
  (\bibinfo{year}{2004}).
  
\bibitem{riccobono_oxidation_2014}
\bibinfo{author}{Riccobono, F.} \emph{et~al.}
\newblock \bibinfo{title}{Oxidation products of biogenic emissions contribute to nucleation of atmospheric particles}.
\newblock \emph{\bibinfo{journal}{Science}}
  \textbf{\bibinfo{volume}{344}}, \bibinfo{pages}{717--721}
  (\bibinfo{year}{2014}).


\end{thebibliography}
\end{document}